\documentclass[export,12pt]{article}

\usepackage[pdftex]{graphicx}
\usepackage{url}
\usepackage[square,sort,comma,numbers]{natbib}
\usepackage{subcaption}
\usepackage{moreverb}
\usepackage{amsmath}
\usepackage[normalem]{ulem}
\usepackage{bm}
\usepackage{adjustbox}
\usepackage[margin=1cm]{caption}
\usepackage{booktabs, siunitx}
\usepackage{mathtools}
\usepackage{afterpage}

\usepackage{tikz}
\usetikzlibrary{arrows}

\begin{document}
\author{Kangyan Liu \footnote{This work was done when he was a Ph.D. Candidate at the University of Connecticut.} \\ 
Motorola solutions 
\\kangyan.liu@motorolasolutions.com  
\and Haim Bar\\ Department of Statistics University of Connecticut,\\ Storrs CT, 06269, USA.}
\title{Large-P Variable Selection in Two-Stage Models}
\maketitle

\abstract{Model selection in the large-P small-N scenario is discussed in the framework of
two-stage models. Two specific models are considered, namely, two-stage least squares (TSLS) involving
instrumental variables (IVs), and mediation models. In both cases, the number of putative variables 
(e.g. instruments or mediators) is large, but only a small subset should be included in the two-stage model.
We use two variable selection methods which are designed for high-dimensional settings, and
compare their performance in terms of their ability to find the true IVs or mediators.
Our approach is demonstrated via simulations and case studies.
}
\section{Introduction}
This paper deals with the problem of variable selection in two-stage models, when the number
of predictors ($P$) in one of the estimation stages is large, possibly greater than the sample size ($N$).
We consider two such models: two-stage least squares (TSLS) involving
instrumental variables (IVs), and mediation.

TSLS is used when some predictor $X$ in a linear model is endogenous, meaning that the assumption
that the errors are not independent of the predictor is violated. If this problem is not
addressed the estimate obtained from ordinary least squares may be biased. A common approach
to dealing with endogeneity is to introduce an IV, $Z$,  a variable which is associated with $X$
but is uncorrelated with the errors. Then, in the first stage we fit an OLS model in which $X$ is
the dependent variable and $Z$ is the independent variable, and in the second stage
we replace $X$ with the predicted version from the first stage.
This can be generalized to a set of endogenous predictors and a set of IVs. The
process of selecting a proper IV in the classical literature is manual and requires careful considerations
and justifications for selecting a specific IV.
Recently,  Belloni et al. \cite{belloni2012sparse} developed a modern approach where the number of proposed IVs is large,
and without a screening step the set of putative IVs is likely over-identified. They showed how variable selection
via the lasso \cite{tibshirani1996regression} can be used to select a small subset of instruments.

Mediation models \cite{Baron1986} are used when the effect of a predictor $X$ on the response $Y$ is indirect.
In the case of so-called complete mediation, the direct effect of $X$ on $Y$ is 0, but
there is a variable $M$ which affects $Y$ and is affected by $X$. We say that the effect of $X$ 
on $Y$ is mediated by $M$.
Mediation analysis is widely applied in psychological science \cite{mackinnon2012introduction}. While the mediation model is built from a case with single mediator and single putative predictor, handing multiple mediators or predictors has become prevalent. \cite{hayes2014statistical} describes approaches to analyze multicategorical independent variables.
The statistical inference is conducted by significance level correction, to account for multiple testing.
The R package `mediation' \cite{tingley2014mediation} uses statistical model-based and design-based approach to implement
a comprehensive causal mediation effects analysis. Although Tingley et al. deal with multiple mediators (\cite{tingley2014mediation} section 6) they assume that the mediators (anxiety and economic harm of immigrants) as well as all the covariates (age, education level, gender, and income) are  relevant in the experiment and no variable selection is conducted.
We consider the case where the number of putative mediators  and/or predictors is large.

The power and accuracy of the variable selection step in two-stage models is of critical importance.
Clearly, failing to detect true predictors (IVs or mediators) has a detrimental effect on the 
second stage. If an IV is not detected, the estimate of the parameter of interest is likely to be biased.
If a mediator is not detected, the analysis will fail to detect an existing indirect association 
between $X$ and $Y$, and when there is complete mediation it means that the relationship
between the two variables will be missed altogether.
What may be less intuitive is the effect of false detection of IVs or mediators. We show that this,
too, has a negative effect on the second-stage estimation.
The detrimental effect of falsely detected predictors was studied in \cite{bar2018} in the context of binary classification
 via logistic regression, support vector machines, classification trees, or random forest. The authors show the
 importance of pre-screening variables when the number of putative predictors is large, but also that if the
 screening step yields too many false predictors, the classifier's accuracy may suffer. Likewise,
 in this paper we show via simulations that inaccurate variable selection may lead to incorrect conclusions in two-stage models.

We introduce the necessary background and notation for TSLS and mediation models, and to
two variable selection methods in Section \ref{sec:background}.
In Section \ref{sec:simulation} we describe simulation results, and in Section \ref{sec:casestudy} we revisit
the `eminent domain' analysis from \cite{belloni2012sparse}. In Section \ref{sec:cancer} we 
demonstrate a novel application of our method in a case where there are thousands of putative mediators (genes)
and we investigate whether any of them mediate the effect of a certain breast cancer subtype (HER2) on 
time to death.
 We conclude with a discussion in Section \ref{sec:discussion}.

\section{Background and Notation}\label{sec:background}
\subsection{Two-Stage Least Squares}
In a linear regression model $Y = X \bm\beta + \bm\epsilon$ it is assumed that the error terms $\epsilon_i$ are independent
and identically distributed as $N(0,\sigma^2)$ and they are uncorrelated with the predictors. In econometrics,
these assumptions are often unreasonable. When the independent variable $X$ and the errors $\bm\epsilon$ are correlated, 
the ordinary least squares (OLS) estimator is not consistent \cite{mkb}. An IV is a variable $Z$, which satisfies $E(\bm\epsilon|Z)=0$ and
$Var(\bm\epsilon | Z)=\sigma^2I$, as well as $plim ~n^{-1}Z'X=\Psi$ and $plim ~n^{-1}Z'Z=\Theta$, both non-singular as
$n\rightarrow\infty$. If $Z$ and $X$ have the same dimensions, the IV estimator of $\beta$ is defined as
$$\bm\beta^*=(Z'X)^{-1}Z'Y$$
and it can be shown to be consistent. However, if the number of columns in $Z$ and $X$ are different, the two-stage
least squares (2SLS or TSLS) approach is used, yielding
$$\bm\beta^{**}=(\hat{X}'\hat{X})^{-1}\hat{X}'Y$$
where $\hat{X}$ is the fitted value of X when it is regressed on $Z$ using OLS:
$$\hat{X}=Z(Z'Z)^{-1}Z'X\,.$$
When choosing an IV, it is important that in addition to being independent of the errors, it is also correlated
with the endogenous predictors, $X$. If that is not the case, or if the correlation is weak, then the
IV is considered weak in the sense that it provides a poor prediction for $X$, which renders the 2SLS estimator
inaccurate \cite{bound1995,wooldridge2008introductory}. 

Two other common estimation methods for handling endogenous variables in linear models
are Fuller's \cite{fuller1977some} and the Limited Information Maximum Likelihood Ratio (LIML) estimator \cite{rivers1988limited}.
We do not elaborate  on these methods here, as it would be outside the scope of our paper, but we will refer
to them in the simulations below.

In this paper we are concerned with situations in which the number of columns in $Z$ is large, and while all
the putative variables (columns) are assumed to be valid instruments, we have to choose a small subset
so that the selected IVs solve the endogeneity problem, but are not over-identified.
Finding valid instruments is an important, and generally difficult task. Choosing a weak instrument may lead to
large inconsistencies in the IV estimates, and choosing an overidentified set of instruments
may cause severe biases in the TSLS estimation procedure, and these biases can be as bad as the ones
from the OLS procedure in the presence of endogenous predictors \cite{bound1995}. This is true even if the sample size is very large.

\subsection{Mediation}\label{sec:mediation}
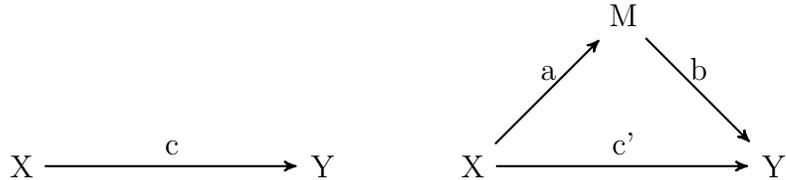
\begin{figure}[!t]
\centering
\begin{tikzpicture}[->,>=stealth',shorten >=1pt,auto,node distance=3cm,
  thick,main node/.style={,fill=white!20,draw,font=\sffamily\small\bfseries}]
\path (-6,0) node(x) {X}
(-2,0) node (y) {Y};
\draw[->] (x) -> node[above] {c} (y);
\hspace{0cm}
\path (0,0) node(x) {X}
(4,0) node (y) {Y}
(2,2) node(m) {M};
\draw[->] (x) -> node[above] {c'} (y);
\draw[->] (x) -> node[above] {a} (m);
\draw[->] (m) -> node[above] {b} (y);
\end{tikzpicture}
\caption{Mediation model. Left: $c$ is the total effect of $X$ on $Y$. Right: $c'$ is the direct effect, $ab$ is the indirect effect.}
\label{fig:mediation}
\end{figure}

Following \cite{Baron1986}, consider two variables with a causal relationship $X\rightarrow Y$.
We say that another variable, $M$, mediates the effect of $X$ on $Y$ if $X\rightarrow M$ and
$X+M\rightarrow Y$. This is described in Figure \ref{fig:mediation}, with paths denoted by lower case letters.
If, after controlling for the effect of $M$ on $Y$, the effect of $X$ is zero, we say that there is complete mediation.

Baron and Kenny \cite{Baron1986} and others describe four steps involved in fitting a mediation model.
In the first step, establish that $X$ is correlated with $Y$ (path $c$).
In the second step, show that $X$ is correlated with $M$ (path $a$). As in the case of IV, this means that we treat $M$ 
as the dependent variable in this step.
In the third step, establish that $M$ is correlated with $Y$ (path $b$), while controlling for $X$.  
In step 4, show that the effect of $X$ on $Y$, while controlling for $M$, is zero (path $c'$).
Since the model has been introduced, there have been discussions as to which of the four steps is essential,
and which is optional. Recent consensus \cite{kenny2014power} states that in order to determine mediation, it is sufficient to perform steps 2 and 3,
and we adopt this approach. So, the fitting procedure consists of the following two steps:
\begin{enumerate}
\item fit $M \sim X$ to test if $a=0$
\item fit $Y \sim M + X$ to test if $b=0$
\end{enumerate}
If neither $a$ nor $b$ are zero, we conclude that the effect of $X$ on $Y$ is mediated (at least in part)
by $M$. In this paper we are concerned with situations in which either the number of independent variables or
the number of mediators (or both) is large.

\subsection{Variable Selection}
The linear model $Y = X \bm\beta + \bm\epsilon$ has no closed form solution when the number of columns in $X$
exceeds the number of rows. For situations like this, which have become quite common in recent years,
new methods have been proposed. In the following section we use two such methods, namely, the lasso 
\cite{tibshirani1996regression} and SEMMS \cite{bar2019}. Both methods yield sparse solutions, but
through different approaches. Before we very briefly describe each method, we want to point out that
many other variable selection methods exist, including variants of the lasso,
such as SCAD \citep{FanLi:2001} and the adaptive LASSO \citep{Zou:2006}, and related methods like 
LARS \citep{Efron:2004,Hesterberg:2010}, or Bayesian 
(Spike and Slab) approaches \cite{Ishwaran:2005,Rockova:2014}. 
The reason we use the lasso is that it is arguably the most popular method,
and because it was used in the seminal paper by Belloni et al. \cite{belloni2012sparse} to derive 
theoretical properties with regard to variable selection in 2SLS models with a  large number of putative IVs.
In the simulations and the case study we can compare our results when using SEMMS with the results in 
\cite{belloni2012sparse}, and the replication study in  \cite{spindler2016lasso}. We use SEMMS because it was shown
in \cite{bar2019} to outperform several competing methods in terms of its power to detect true predictors,
while maintaining a low number of false positives. In particular, it was shown to perform better than competing
methods in cases in which the predictors were highly correlated.

The lasso estimator is obtained by optimizing the following constrained optimization problem:
$$\hat{\bm\beta}\in \arg\min_{\bm\beta} \sum_{i=1}^n(y_i-x_i'\bm\beta)^2 + \lambda\|\bm\beta\|_1$$
where $\|\bm\beta\|_1 = \sum_{j=1}^P|\beta_j|$.
In other words, minimize the sum of squared errors subject to an $\ell_1$ penalty on the predictors.
The tuning parameter $\lambda$ may be selected by the user, but more commonly, it is estimated 
by the software via cross-validation. There are multiple implementations of the lasso estimator, and in the simulations below
we chose to use the ncvreg package \cite{ncvreg}.

As an alternative to the penalty-based approach of the lasso, we also use
the empirical Bayes approach of \cite{bar2019}.
The variable selection task is cast as a classification problem by expressing the relationship between
the expected value of the response and the predictors as follows:
$$E(y_{i})= \sum_{k=1}^{P}x_{ik}\gamma _{k}u_{k}\,,$$
such that, conditional on the linear predictors, the $N$ responses are i.i.d. normal realizations.
The latent variables are assumed to follow a multinomial distribution,
 $\gamma _{k} \stackrel{iid}{\sim} multinomial\left(-1,0,1; p_L,p_0,p_R\right)$, and the terms $u_k$ are
 treated as random coefficients, such that $u_{k} \stackrel{iid}{\sim} N\left( \mu ,\sigma ^{2}\right)$,
independently of $\gamma _{k}$.
This amounts to  $\gamma _{k}u_{k}$ belonging to one of three components,  $C_L$, $C_0$, and $C_R$,
corresponding to $\gamma _{k}=-1,0,1$. Therefore, with this mixture model approach, variables for which 
$\gamma_k\ne0$ are selected to be included in the linear regression model.

Estimation of the model parameters is done via a Generalized Alternating Minimization (GAM, \cite{Gunawardana:2005})
 algorithm which is a generalization of the EM algorithm \cite{Dempster1977}, and like its predecessor, is computationally efficient
 and guaranteed to converge.

We point out that both lasso and SEMMS can be used to perform variable selection in the 
generalized linear model (GLM) framework, but here we focus on the normal setting only.
With SEMMS the user can specify if any predictors must always be included in the model (`locked in'
variables), while in current implementations of the lasso there is no such option. This is especially useful in
two-stage procedures, such as the mediation estimation procedure we described above. 

\section{Simulations}\label{sec:simulation}

\subsection{Instrumental Variables}
In this section, we follow the so-called `cut-off' scenario in \cite{belloni2012sparse} and \cite{spindler2016lasso}
 to simulate data with a large 
number of putative IVs, according to the following specifications.
With  the cut-off design we can ensure that the model for the first stage is sparse, meaning
that the number of instruments associated with the
independent variables is small. 
Denote $\bm\gamma=C\cdot(\bm{1}_L,\bm{0}_{P-L})^{'}$ where $P$ is the total number of putative instruments,
and $L$ is the number of instruments which are actually correlated with the dependent variables.
The constant, $C$, is derived below.
Let the instruments $Z$ follow a multivariate Gaussian distribution 
\begin{eqnarray}
Z&\sim& N_P(\mathbf{0},\Sigma)\mbox{ such that }\sigma_{ij}=0.5^{|i-j|}\,.
\end{eqnarray}
Then, we define
$$\sigma_v^2=1-\bm\gamma^{'}\Sigma\bm\gamma$$
and the data are generated according to the following hierarchical equations:
\begin{eqnarray}
(e_i,v_i)&\sim& N 
\begin{bmatrix}
0,&
\begin{pmatrix}
1 & 0.6 \\
0.6 & \sigma_v^2
\end{pmatrix}
\end{bmatrix}\\
x_i &=& \mathbf{z}_i \bm\gamma + v_i\\
y_i &=& x_i \beta + e_i\,.
\end{eqnarray}
Note that in general the relationship between regressors and the instruments need not be linear, but it is
a popular and convenient model. Also, to simplify things (and to be consistent with \cite{spindler2016lasso})
we assume that there is a single independent variable, $x_i$. 
Also note that
the condition $\sigma_v^2=1-\bm\gamma^{'}\Sigma\bm\gamma$ ensures that the unconditional variance of $x_i$ equals one.
The fact that the two error terms are correlated induces endogeneity.

The constant $C$ in the definition of $\bm\gamma$ is calculated by solving the equation involving
the so-called concentration parameter $\mu^2$ which controls the scale of the second-stage coefficients:
$$\mu^2=\frac{P\cdot C^2(\bm{1}_L,\bm{0}_{P-L})^{'}\Sigma(\bm{1}_L,\bm{0}_{P-L})}{1-C^2(\bm{1}_L,\bm{0}_{P-L})^{'}\Sigma(\bm{1}_L,\bm{0}_{P-L})}\,.$$ 

As in \cite{spindler2016lasso}, we set $\beta=1$. We vary three parameters in the model:
the sample size $n=100,250$, 
the number of true predictors for $x$, $L=5,10,20$, and the
concentration parameter $\mu^2=30,180$. 
One way in which our simulations differ from \cite{spindler2016lasso} is in the magnitude of 
$P$, the number of putative instruments.
Spindler \cite{spindler2016lasso} used $P=100$, whereas we used $P=500$ which exceeds even the largest
of the sample sizes in our simulation. Combined with the fact that $L\ll P$, this makes the 
variable selection task in stage 1, considerably more challenging.

For each simulation configuration, we apply the lasso and SEMMS in the first stage,
and evaluate their performance in terms of the following criteria:
\begin{itemize}
 \item The number of times no instruments were found, $N(0)$.
 \item The bias in the estimation of the parameter of interest, $\beta$.
 \item The mean absolute deviation (MAD) of the estimate for $\beta$.
 \item The number of true positive instruments (out of $L$).
 \item The number of false positive instruments (out of $P-L$).
 \item The p-value from the first stage.
 \item The coverage probability of $\beta$.
\end{itemize}

Table \ref{tslssim} shows the results reported as averages of 100 repetitions of each simulation scenario. 
The variable selection process applies only to the first stage and was performed 
with two R packages with the following settings.
For Lasso we used the ncvreg package \cite{ncvreg} with the Gaussian family and lasso penalty settings,
and 10-fold cross-validation for parameter tuning.
SEMMS \cite{SEMMS} was used with the default settings. In the simulations we performed the
two-stage estimation using the classical 2SLS approach, as well as the preferred Fuller 
\cite{fuller1977some} and limited information maximum likelihood ratio (LIML, \cite{rivers1988limited}) methods.

When the concentration parameter is set at the low level, $\mu^2=30$, lasso fails to detect any
of the true instruments 1-4 times for $L=5$, 11-14 for $L=10$, and 20-22 times for $L=20$. In the
high $\mu^2$ setting the lasso always detects at least one true instrument.
SEMMS, on the other hand failed to detect any true instrument just once in 1,200 simulations (in the
$\mu^2=30$ and $n=250$ setting.)

In terms of the accuracy of $\hat\beta$, SEMMS yields smaller bias and mean absolute deviation
 in all scenarios,
compared with the lasso. The bias produced when the lasso is used in the first stage can be quite substantial.
For example, recall that $\beta=1$, and notice that when the true number of predictors is 20, 
the concentration parameter is 180, and the sample
size is 250, the bias is 0.21 for Fuller and LIML, and 0.29 for TSLS (and only 0.09 and 0.11, respectively,
when SEMMS is used).  
When the instruments are strong, SEMMS has a clear advantage over the lasso. For example,
using LIML in the second stage, in the case of $L=10$ and
$n=100$ the bias and MAD following variable selection with SEMMS were 0.04 and 0.1, respectively,
versus 0.19 and 0.21, respectively, with the lasso.

Comparing true/false positive rates, it appears that SEMMS is slightly more conservative, yielding fewer
true positives on average. However, keep in mind that in the specific simulation settings the true
instruments are highly correlated, so the ones actually detected by SEMMS provide sufficient prediction power.
Furthermore, SEMMS automatically locks-out variables which are strongly correlated with selected variables,
so as to eliminate multicolinearity. Table \ref{tslssim} does not include those `locked-out' variables in the TP
columns, thus the true number of TP found by SEMMS is actually higher than reported in the table.
In terms of false positives, SEMMS has a clear advantage over the lasso. For example, in the case where 
$L=5$, $\mu^2=180$, and $n=100$, lasso has on average, 14 false positives, compared with only 0.5
for SEMMS. Across all scenarios, the false positive for the lasso ranges between 10.77 and 26.39,
 whereas SEMMS yields only 0.08-1.77 false positives.
These results are consistent with the ones reported in \cite{bar2019}.
It appears that the inclusion of unrelated instruments lead to higher bias.

The p-value columns in Table \ref{tslssim} show the p-value obtained when fitting the IV model with
the null hypothesis $H_0:\beta=1$. In all cases, SEMMS yielded a higher p-value than lasso, and
never falsely rejected the null hypothesis. With lasso, on the other hand,  if we were to use the TSLS approach,
we would falsely reject $H_0$ when $L=20$ (and $L=10$ with $\mu^2=30$.)
The CP columns show the 95\% coverage probability (the proportion 
of simulations for which the confidence interval contained the true value of $\beta$, when using a significance
level of 0.05.)
In all cases SEMMS yielded a higher coverage probability than the lasso, and in the case of strong instruments
SEMMS achieved a coverage probabilities of 0.92-0.94 when $L=5$, 0.91-0.92 when $L=10$, and 0.84-0.86
when $L=20$ (using Fuller or LIML). In contrast, the CP of the lasso were  0.61-0.68 when $L=5$,
0.52 when $L=10$, and only 0.32-0.41 when $L=20$.

\subsection{Mediation}
We simulated a dataset with complete mediation per Figure \ref{fig:mediation}, with $N=100$ observations:
\begin{eqnarray*}
M &=& 1 + \beta_1 X + \epsilon_M \text{ (effect a)}\\
Y &=& 1 + \beta_2 M + \epsilon_Y \text{ (effect b)} \,,
\end{eqnarray*}
where both $\epsilon_M$ and $\epsilon_Y$ are i.i.d. from a normal distribution, $N(0,0.2)$, and
$(a,b)\in\left\{(3,1), (\sqrt{3},\sqrt{3}), (1,3)\right\}$. 
We consider two main scenarios -- one with multiple putative mediators, $M_j$, $j=2,\ldots,500$ and one main effect, $X$; 
and the other with multiple main effects, $X_j$,  $j=2,\ldots,500$ and one mediator, $M$.  We use the notation $X_1$ and $M_1$
to refer to the original variables from the data generation model. To simplify the notation, we denote
$V_j$ as the predictors (either $M_j$ or $X_j$) and $Z$ as the response (respectively, $Y$ and $M$.) 
In both scenarios the additional 499 variables are created using one of the following settings:
\begin{enumerate}
\item $V_j$, $j=2,\ldots,500$ are i.i.d. from a standard uniform distribution and all are uncorrelated
 with the response, $Z$.\label{iidcase}
\item Similar to \ref{iidcase}, but $V_1,V_2,\ldots,V_{10}$ have an AR(1) correlation structure, with $\rho=0.7$.\\
$V_j = V_{j-1} + N(0,0.3^2)$ for $j=2,\ldots,10$\label{ar1case}
\item Similar to \ref{ar1case}, but each of $V_2,\ldots,V_{10}$ is set to have the same 
correlation ($\rho=0.7$) with $V_1$.\\
$V_j = V_{1} + N(0,0.3^2)$ for $j=2,\ldots,10$\label{blockcase}
\item Similar to \ref{blockcase}, but $V_2,V_3,V_4$ are positively correlated with $V_1$ and $V_5,\ldots,V_{10}$
are negatively correlated with $V_1$.\\
$V_j = V_{1} + N(0,0.3^2)$ for $j=2,3,4$\\
$V_j = -V_{1} + N(0,0.3^2)$ for $j=5,\ldots,10$\label{mixedblockcase}
\end{enumerate}
Each simulation was repeated $B=500$ times.

We check whether the variable selection methods considered in the paper, namely SEMMS and lasso,
are able to find the true underlying complete-mediation model, and whether the
true effects are estimated accurately.
In case \#\ref{iidcase} there is only one true main effect (or mediator) while in the other three cases, because of the
strong correlations, any of the first 10 predictors (mediators) might be considered valid, if found by a variable selection method.
Let $V_T$ denote the set of valid (true positive) predictors, and let $S$ be the set of predictors selected by the variable selection
algorithm.
Using $\alpha=0.05$, we summarize the results for both scenarios (multiple mediators or multiple main effects) in terms of
\begin{itemize}
 \item The number of times the true predictor or mediator were not found, $N(0)=\#\left\{V_1\notin S\right\}$.
 \item The average number of true positives, $TP=\#\left\{S\cap V_T\right\}/B$.
 \item The average number of false positives, $FP=\#\left\{S\cap \overline{V_T}\right\}/B$.
  \item For the following items, results are counted only when the  true predictor or true mediator is found and 
  the model $Y=\beta_0+ \beta_1 X+ \beta_2 M$ is used per Step 2 in Section \ref{sec:mediation}.
  \begin{itemize}
   \item The percentage of times the mediation effect ($b$, per Figure \ref{fig:mediation}) was correctly identified as significant, $\beta_2\ne0$ .
 \item The percentage of times the direct effect ($c'$, per Figure \ref{fig:mediation}) was incorrectly 
 identified as significant, $\beta_1 \ne0$. 
 \item The bias in the estimation of $\beta_2$.
  \item The mean absolute deviation in the estimation of $\beta_2$.
   \item The  coverage probability in the estimation of $\beta_2$.
\end{itemize}
\end{itemize}
The results are summarized in Table \ref{medisim}. In all the scenarios, the total effect of $X$ on $Y$ is 3.
It can be seen that as the effect size of $X$ on $M$ gets smaller, the lasso has more false positives,
while SEMMS has an average of nearly zero false positives in all 6 scenarios. 
Note that the reported results consist of averages, and in some simulations the total number of false positives detected
by the lasso was as high as 47.
SEMMS also achieves decent true positive rates, while
the lasso had a total of 168 cases where $X_1$ (or $M_1$) was not detected. In the three configurations in which
the predictor (or mediator) is correlated with 9 other variables the lasso detects at most 4.63 out of 10 true positives.

Both algorithms detect the significant relationship between $X$ and $M$ with high probability. SEMMS achieves
100\% accuracy in this regard, while the lasso achieved 89.8\% and 89.9\% in the scenario with multiple mediators 
with $\beta_2=1$ and the correlation structure between the mediators per cases \ref{blockcase} and \ref{mixedblockcase}
above. 
 Both methods yield a small bias for $b$, but again SEMMS is less sensitive to the scenario settings. The mean absolute deviation was approximately 0.084 in all scenarios when using SEMMS, which was almost always smaller than the MAD
 of $\beta_2$ when using lasso (e.g., with multiple mediators and $\beta_1=1$, $\beta_2=3$, the MAD is 0.246.)
Similarly SEMMS yielded approximately the same coverage probability of $\beta_2$ (with $\alpha=0.05$) in all scenarios, and very close to
the desired value of 0.95 (0.943-0.945). The lasso yielded generally good results, but in some cases the coverage probability
was slightly lower (e.g., 0.933 in the multiple M scenario with $\beta_1=1$, $\beta_2=3$).

To address a more challenging situation, we further modify the simulation in the multiple X case with setting 4
so that $\beta_1 = \beta_2 = 0.5$, and $\rho$ is increased to 0.9 from 0.7. 
In this case, the bias is small (-0.0155 with the lasso vs. -0.00547 with SEMMS). However,
the lasso detects b as significant only $79.8\%$  of the time, while SEMMS detects a significant b nearly 
100\% of the time.

We also note that in the multiple $X$s scenario the variable selection procedure was equivalent for both methods (that
is, we simply used $M$ as the response, and $X_j$ as the putative predictors.)
However, in the multiple $M$s scenario, SEMMS allows to have `locked in' variables, and thus it is possible to
fit the mediation model, $Y\sim X+\sum_jM_j$, but that is not the case with the lasso. We considered three procedures
which would allow for a fair comparison between the two methods. The first two were to either simply consider $X$ as another
 putative  variable, or to only use the $M_j$s. However, because we simulated complete mediation this seemed inappropriate
 because in mediation models the effects of the mediators are estimated conditionally on $X$ (so, $X$ would have to be considered
 as `locked in' and not subject to selection). Hence, we chose the third option
 which involved treating $X$ as the response and $M_j$ as the putative variables. This is reasonable in this simulation, but conceptually
 contrary to the inherent assumption of causality.

\section{Examples}
\subsection{Instrumental Variables -- The Eminent Domain Study}\label{sec:casestudy}

We illustrate our approach with the `eminent domain' example from \cite{belloni2012sparse}.
The objective of the study is to find whether rulings made by federal appellate courts in eminent
domain cases have an impact on economic outcomes.  
Eminent domain cases involve law suits by individuals whose land has been taken as part of the government's decision
to exercise the `takings law'. We focus on one particular outcome, namely, the Case-Shiller index
 for home prices.
The structural model used in \cite{belloni2012sparse} is
$$y_{ct}=\alpha_c+\alpha_t+\gamma_ct+\beta\cdot TakingsLaw_{ct}+W_{ct}'+\epsilon_{ct}$$
where $y_{ct}$ is the log(Case-Shiller) outcome for circuit $c$ at time $t$, 
$TakingsLaw_{ct}$ is the number of \textbf{pro-plaintiff} decisions in circuit $c$ at time $t$,
and $W_{ct}$ are the judicial pool characteristics used by \cite{belloni2012sparse}, which include: 
gender, race, religion, political affiliation, public/private school education
where the judge obtained his/her BA and/or JD, whether the judge was elevated from a district court,
and interactions of these variables. 
The parameter of interest is $\beta$, which is interpreted as the effect of an additional decision
in favor of a plaintiff (and against the government's decision) on the Case-Shiller index.
Belloni et al. \cite{belloni2012sparse} make the case that the court's ruling on takings law, and the
government's taking of lands may be endogenous. They propose to use the panel characteristics
as instruments because the panels (and hence, their demographics) are randomly assigned to cases,
and are very unlikely to be related to economic outcomes such as home prices. While the argument
for using these characteristics as instruments is compelling, the problem is that one has to choose 
a subset which is both a strong predictor for the TakingsLaw variable, and not over-identified.

We use SEMMS as the variable selection method in the first stage of the eminent domain 
data for the analysis where log(Case-Shiller) is the second-stage dependent variable.
The sample size in this data set is $n=183$ and the number of putative instruments is
$p=147$. 
We use the `greedy' version of the algorithm, meaning that in each iteration we select 
the variable whose addition to, or removal from the model yields the largest increase
in the log-likelihood. To prevent multicollinearity we use the pairwise
absolute correlation threshold of 0.7 between predictors.
SEMMS picks the following five variables as instruments:
\begin{itemize}
\item $3x\_dem$ the number of  panels consisting of three democrats
\item  $1x\_jd\_public^2$ squared number of panels with at least one judge who received his/her JD in a public school
\item  $3x\_elev$  number of panels in which all three judges were elevated from a district court
\item  $1x\_female.*1x\_catholic$ an interaction between the number of panels with at least one female and at least one Catholic judge
\item $2x\_female.*1x\_ba\_public$ an interaction between the number of panels with at least two female judges
and at least one judge who received his/her BA from a public school.
\end{itemize}
All these variables are found to be highly significant in the first stage estimation.
The first-stage regression yields an F-statistic of 7.4 on 5 and 177 degrees of freedom
(p-value=2.52e-06), suggesting that the selected variables are strong instruments
for the TakingsLaw variable.
The Sargan test statistic is 5.75 with 4 degrees of freedom
(p-value=0.22) so there is no evidence of over-indetification by this set of IVs.
Both the Anderson-Rubin and the conditional likelihood ratio tests
lead to the conclusion that, at the 5\% level, the effect of TakingsLaw on 
the log(Case-Shiller) variable is not significant
(p-values 0.098 and 0.065, respectively). The t-values from TSLS, Fuller, and LIML
are 1.726, 1.816, and 1.838, respectively. When we use the heteroscedastic-robust standard errors option in the
ivmodel package \cite{ivmodel}, the corresponding t-values are 
1.502, 1.425, and 1.406, respectively.

We compare our results with the ones in \cite{belloni2012sparse}, where the lasso was used.
Two variables were selected by the lasso ($1x\_jd\_public$ and $1x\_jd\_public^2$)
with a first-stage p-value of 0.0004, and a Sargan test statistic of 0.04 (df=1, p-value=0.84).
The Anderson-Rubin test gives a p-value of 0.043, and the conditional likelihood ratio test
gives p=0.014, suggesting that, at the 5\% level the effect of TakingsLaw on 
the log(Case-Shiller) variable is significant.
The TSLS, Fuller, and LIML methods give similar t-values of 2.157, 2.13, and 2.158
(with p=0.032, 0.035, and 0.032), respectively.
With the heteroscedastic-robust standard errors option, the corresponding t-values are 1.967, 2.046, and 1.964
(with p=0.051, 0.042, and 0.051), respectively.

\begin{figure}
 \centering
\includegraphics[scale=0.7]{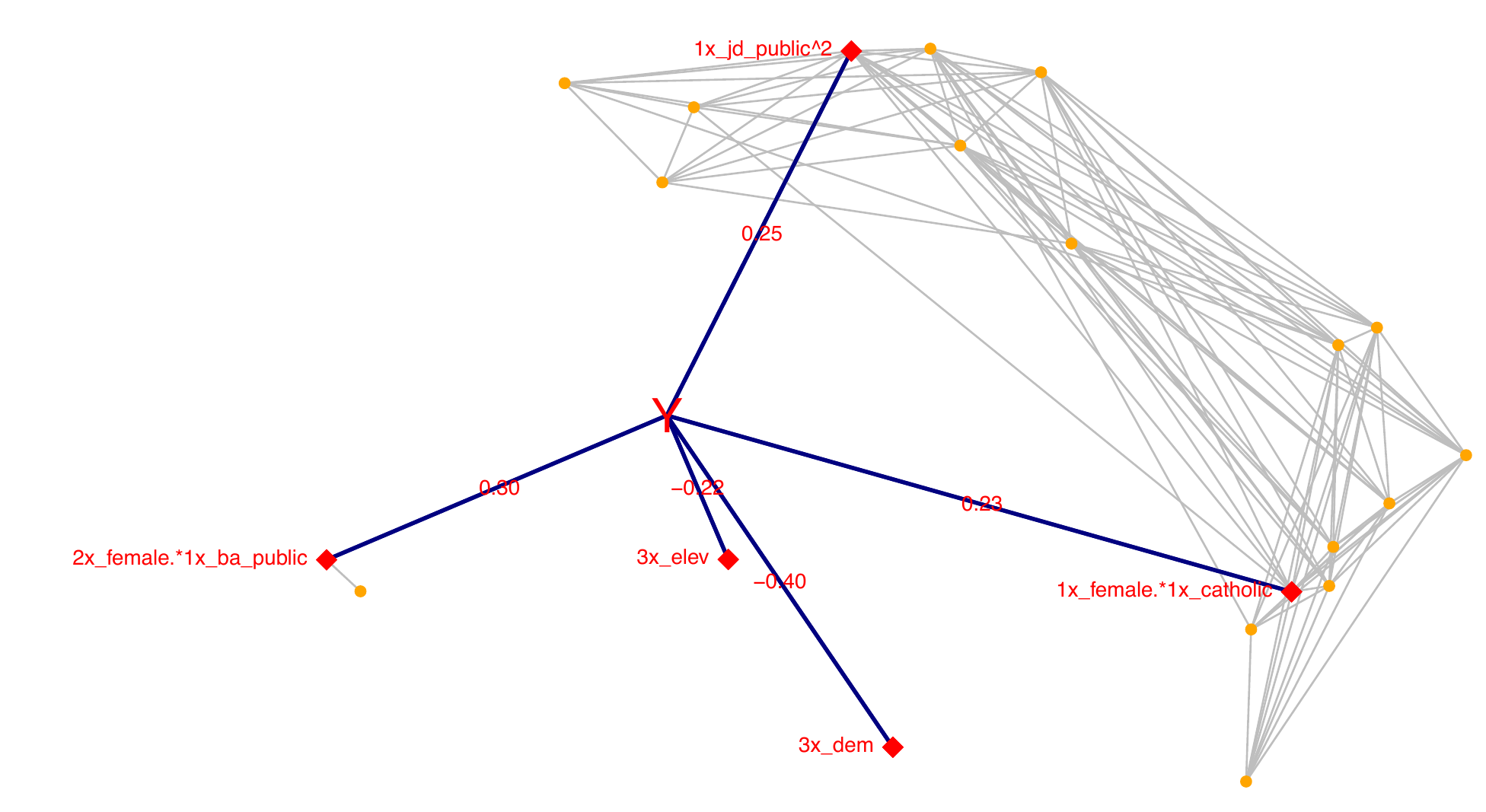} 
 \caption{A graphical representation of the model selected by SEMMS
 for the first stage regression in the analysis of the eminent domain data (with log(Case-Shiller)
 as the dependent variable in the second stage.)}\label{figure:EDsemms}
\end{figure}

Figure \ref{figure:EDsemms} shows a graphical representation of the model selected for the 
first stage by SEMMS. The five selected predictors appear as red diamonds and are connected to
the dependent variable ($Y$=TakingsLaw) with a dark blue edge. 
The orange dots represent putative variables not selected to be included in the model,
but that were found to be highly correlated ($|r|>0.7$) with at least one selected
variable (depicted by light grey lines). It is clear that two of the selected variables  
($1x\_jd\_public^2$   and  $1x\_female.*1x\_catholic$) are part of a rather large interconnected
network of predictors, each of which may be useful as a strong instrument (among them we find
$1x\_jd\_public$, which was selected by the lasso). However,
including all variables in this network is likely to result in over-identification.
It can also be seen by the blue edge labels that panels with three democrats or with three judges
who were elevated from a district court tended to rule against the plaintiff (and in favor of the state),
as indicated by the negative effect on the first-stage response. The other three
instruments have a positive effect, which means that panels with these characteristics tended to rule for the
plaintiffs.

In conclusion, IVs selected by SEMMS provide stronger relationship to the dependent variable in the first
stage (namely, TakingsLaw) than the lasso, without risking over-identification. Taking into account 
the correlations between putative predictors is also an advantage of using SEMMS in the first stage, since it reveals
important relationships between putative IVs (so there may be multiple combinations of IVs that can be 
used in the first stage), and prevents multicolinearity at the same time. 
Our results suggest that, in contrast to \cite{belloni2012sparse}, there is no causal effect of decisions in 
favor of plaintiffs in eminent domain cases on property values as measured by the Case-Shiller index. 
If one accounts for heteroscedasticity of the errors, the results with the IVs selected by the lasso 
in  \cite{belloni2012sparse} also suggest that the causal relationship may not be significant (using the LIML estimation approach).
The results from the first stage estimation with SEMMS are interesting not only in the sense that the 
potential IVs are interconnected as mentioned earlier, but also that cases which were assigned \textit{all} three
judges who were all democrats, or \textit{all} have been elevated from a district court tended to rule against the plaintiffs.

\subsection{Mediation -- Breast Cancer Study}\label{sec:cancer}
We illustrate our approach for mediation models in the case where the number of
putative mediators is very large. We use the ACES dataset  \cite{Allahyar2019,ACES} which combines
twelve breast cancer studies, with a total of 1,616 samples.
We focus on the following question: among breast cancer patients who
did not survive, was there a difference between women with the HER2 subtype and other patients
in terms of the time to death, and if so, was this difference mediated by certain genes?


There are several breast cancer subtypes, and the HER2-enriched type tends to grow faster 
than other types, and generally has a worse prognosis, although there are effective targeted therapies.
From the ACES dataset we extracted of subset of patients who died, and were in the age range
of 35-65. The total number of patients was 220 (from the `Miller' and `Desmedt-June07' studies).
We considered $Y=$ `time to death' to be the response, and the HER2 (binary) status to be
the predictor, $X$. We considered the expression levels of 12,750 genes as putative
mediators. The causality diagram in Figure \ref{fig:mediation} is plausible, since a mutation
may alter the expression levels of certain genes, and these, in turn may have an effect on
phenotypes. It is also conceivable that the effect of a mutation on a phenotype is mediated (partially
or completely) by some genes.

We apply SEMMS to the model $Y_i \sim \sum_{j=1}^{12750}M_{ij} + X_i$ so that $X$ is 
always included in the model, and SEMMS selects the significant $M_j$'s. 
The algorithm yields a single mediator, the Adrenomedullin gene (AMD, Entrez 133).
The edge labeled $a$ in Figure \ref{fig:mediation} is significant, and the effect of the HER2
indictor on the expression level of the AMD gene is positive (0.677, p=0.002).
Controlling for $X$, the indirect effect of AMD on time to death is negative 
and significant (-432, p=2.32e-05). In the second stage model (establishing whether $b=0$ in
Figure \ref{fig:mediation}) we get that the direct effect of HER2 on time to death
is not significant (p=0.7), which suggests complete mediation.
Thus, terminal breast cancer patients with the HER2 subtype are more prone to have elevated
AMD expression levels, and a unit increase in normalized AMD levels corresponds to
432 days shorter life span, on average. 

The potentially mediating gene has been linked to accelerated breast cancer bone metastasis 
\cite{Siclari}, to acute episodes of the systemic capillary leak syndrome \cite{Xie}, and 
\cite{Zhou} reported that the AMD gene promotes intrahepatic cholangiocellular carcinoma metastasis.
Identifying AMD as a potential mediator between being classified as a HER2 subtype of breast cancer
and the patient's higher risk for shorter life span could have important therapeutic implications,
especially since the relationship between the subtype and the outcome is completely mediated
by AMD, and thus could not be revealed with models involving only direct effects.

Note that a mediation model which includes the treatment method as a predictor in addition to the HER2
subtype, would be of interest. However, in the two breast cancer studies used in our analysis 
there is scarce information about the treatment. Out of 220 patients, the treatment field is missing
for 186. Twenty eight patients received no treatment, and the remaining six received tamoxifen.

\section{Discussion}\label{sec:discussion}
Two-stage models are commonly used in economics, psychology, and other fields, as they provide 
a convenient and rich framework to model causality and account for possible endogeneity.
As is the case with linear regression and analysis of variance, the original applications of two-stage methods
involved a carefully chosen set of predictors and instruments/mediators. However, with advancements in technology
and the abundance of data it has become quite natural to  consider situations which involve many predictors and instruments/mediators,
possibly in numbers that exceed the sample size. For these situations, a screening step involving a variable selection 
algorithm is essential.
We presented two such cases -- one from the economics literature involving house prices, and one from genomics, involving 
survival times of breast cancer patients.

We showed here that the accuracy of the screening step is important in two-stage models. Even if the bias of the
estimated parameters is small, the conclusion from the two-stage model may be wrong if the screening step fails to
detect all the true predictors, instruments, or mediators, but also if it detects too many false ones. For example, in some
cases in our simulation of a mediation model with multiple predictors we saw that with one variable selection method
(lasso) we may fail to detect the mediation effect 20\% of the time, while with another (the empirical Bayes method called SEMMS)
the mediation effect was detected as significant nearly 100\% of the time. Similarly, in our IV simulations we saw that the lasso
tends to yield a more biased estimate for the first-stage parameter, and have a lower coverage probability compared with SEMMS.

In our case studies it was evident that some of the putative variables were highly correlated. This presents a challenge to many
variable selection methods, including the lasso, but more importantly, when this is the case, it is important for 
the researcher to know that the predictors are not independent. This means that there is not necessarily one `correct'
model that the algorithm has to find. Rather, there can be a complicated combination of predictors which have to
be considered, and there can be even competing models involving different sets of predictors.

We intend to investigate an extension of the approach discussed here, to similar but more complicated models, involving more
than two stages. For example, we will consider an extension to the structural equation modeling (SEM) and possibly to
deep-learning models.

\bibliographystyle{abbrv}
\bibliography{TSVSmodels}


\begin{table}%
	\clearpage
	\thispagestyle{empty}
	\centering
	\scriptsize
	\setlength\tabcolsep{2.5pt}  
	\begin{adjustbox}{ addcode=
			{\begin{minipage}{\width}}
				{\caption{Results are based on 100 simulation replications and 500 instruments. Column labels have two groups (SEMMS and Lasso) with three methods (TSLS, Fuller and LIML) for each group; in cases where no instrument is selected, values are excluded from average. We report the number of
						replications in which   no instrument is chosen $N(0)$, mean of bias, mean absolute deviation (MAD), true positive and false positive for selection. P-values and coverage probability (CP) of the tests on first stage are also reported.}\label{tslssim}
			\end{minipage}},
			height=7cm,width=\textheight,
			angle=90}
		\begin{tabular}{r|rrrrrrr|rrrrrrr|rrrrrrr}
			\hline
			\multicolumn{7}{c}{L=5}&\multicolumn{7}{c}{L=10}&\multicolumn{7}{c}{L=20}\\
			\hline
			& N(0) & Bias & MAD & TP & FP & p-value&CP& N(0) & Bias & MAD & TP & FP & p-value&CP& N(0) & Bias & MAD & TP & FP & p-value&CP \\ 
			\hline
			\multicolumn{22}{c}{A. Concentration parameter = 30, n=100}\\
			\hline
			SEMMS & 0.00 & 0.31 & 0.31 & 1.75 & 1.77 & 0.18 & 0.47 & 0.00 & 0.36 & 0.36 & 1.86 & 1.97 & 0.11 & 0.35 & 0.00 & 0.40 & 0.40 & 1.77 & 2.11 & 0.10 & 0.32 \\ 
			SEMMS-F & 0.00 & 0.28 & 0.29 & 1.75 & 1.77 & 0.22 & 0.60 & 0.00 & 0.34 & 0.35 & 1.86 & 1.97 & 0.14 & 0.42 & 0.00 & 0.39 & 0.39 & 1.77 & 2.11 & 0.11 & 0.40 \\ 
			SEMMS-L & 0.00 & 0.28 & 0.29 & 1.75 & 1.77 & 0.23 & 0.61 & 0.00 & 0.34 & 0.34 & 1.86 & 1.97 & 0.16 & 0.44 & 0.00 & 0.38 & 0.39 & 1.77 & 2.11 & 0.12 & 0.40 \\ 
			Lasso(CV) & 4.00 & 0.37 & 0.39 & 2.75 & 13.94 & 0.09 & 0.23 & 14.00 & 0.42 & 0.42 & 3.26 & 13.49 & 0.07 & 0.17 & 20.00 & 0.44 & 0.44 & 4.33 & 15.11 & 0.05 & 0.13 \\ 
			Lasso(CV)-F & 4.00 & 0.35 & 0.37 & 2.75 & 13.94 & 0.11 & 0.29 & 14.00 & 0.40 & 0.40 & 3.26 & 13.49 & 0.08 & 0.18 & 20.00 & 0.42 & 0.42 & 4.33 & 15.11 & 0.07 & 0.14 \\ 
			Lasso(CV)-L & 4.00 & 0.34 & 0.37 & 2.75 & 13.94 & 0.12 & 0.30 & 14.00 & 0.40 & 0.40 & 3.26 & 13.49 & 0.09 & 0.18 & 20.00 & 0.41 & 0.42 & 4.33 & 15.11 & 0.07 & 0.14 \\ 
			\hline
			\multicolumn{22}{c}{B. Concentration parameter = 180, n=100}\\
			\hline
			SEMMS & 0.00 & 0.04 & 0.10 & 4.44 & 0.50 & 0.50 & 0.95 & 0.00 & 0.07 & 0.11 & 5.64 & 0.19 & 0.47 & 0.90 & 0.00 & 0.11 & 0.14 & 5.50 & 0.26 & 0.42 & 0.84 \\ 
			SEMMS-F & 0.00 & 0.02 & 0.10 & 4.44 & 0.50 & 0.53 & 0.94 & 0.00 & 0.05 & 0.10 & 5.64 & 0.19 & 0.49 & 0.92 & 0.00 & 0.09 & 0.13 & 5.50 & 0.26 & 0.44 & 0.85 \\ 
			SEMMS-L & 0.00 & 0.02 & 0.10 & 4.44 & 0.50 & 0.53 & 0.94 & 0.00 & 0.04 & 0.10 & 5.64 & 0.19 & 0.50 & 0.92 & 0.00 & 0.08 & 0.13 & 5.50 & 0.26 & 0.45 & 0.86 \\ 
			Lasso(CV) & 0.00 & 0.20 & 0.21 & 4.89 & 13.98 & 0.21 & 0.50 & 0.00 & 0.24 & 0.25 & 8.58 & 22.01 & 0.11 & 0.35 & 0.00 & 0.27 & 0.28 & 13.35 & 26.39 & 0.08 & 0.28 \\ 
			Lasso(CV)-F & 0.00 & 0.15 & 0.17 & 4.89 & 13.98 & 0.30 & 0.68 & 0.00 & 0.19 & 0.21 & 8.58 & 22.01 & 0.20 & 0.51 & 0.00 & 0.22 & 0.23 & 13.35 & 26.39 & 0.16 & 0.40 \\ 
			Lasso(CV)-L & 0.00 & 0.14 & 0.17 & 4.89 & 13.98 & 0.30 & 0.68 & 0.00 & 0.19 & 0.21 & 8.58 & 22.01 & 0.20 & 0.52 & 0.00 & 0.22 & 0.23 & 13.35 & 26.39 & 0.17 & 0.41 \\ 
			\hline
			\hline
			\multicolumn{22}{c}{C. Concentration parameter = 30, n=250}\\
			\hline
			SEMMS & 0.00 & 0.29 & 0.29 & 1.96 & 1.90 & 0.18 & 0.37 & 1.00 & 0.31 & 0.32 & 2.12 & 2.18 & 0.13 & 0.41 & 0.00 & 0.37 & 0.37 & 2.20 & 2.19 & 0.09 & 0.31 \\ 
			SEMMS-F & 0.00 & 0.26 & 0.27 & 1.96 & 1.90 & 0.22 & 0.47 & 1.00 & 0.29 & 0.30 & 2.12 & 2.18 & 0.16 & 0.45 & 0.00 & 0.36 & 0.36 & 2.20 & 2.19 & 0.11 & 0.33 \\ 
			SEMMS-L & 0.00 & 0.25 & 0.27 & 1.96 & 1.90 & 0.22 & 0.49 & 1.00 & 0.28 & 0.29 & 2.12 & 2.18 & 0.18 & 0.47 & 0.00 & 0.35 & 0.35 & 2.20 & 2.19 & 0.12 & 0.33 \\ 
			Lasso(CV) & 1.00 & 0.36 & 0.37 & 3.01 & 10.77 & 0.08 & 0.24 & 11.00 & 0.40 & 0.40 & 4.07 & 15.43 & 0.05 & 0.14 & 22.00 & 0.42 & 0.42 & 5.37 & 15.60 & 0.03 & 0.07 \\ 
			Lasso(CV)-F & 1.00 & 0.32 & 0.34 & 3.01 & 10.77 & 0.12 & 0.31 & 11.00 & 0.36 & 0.37 & 4.07 & 15.43 & 0.09 & 0.19 & 22.00 & 0.39 & 0.39 & 5.37 & 15.60 & 0.05 & 0.14 \\ 
			Lasso(CV)-L & 1.00 & 0.32 & 0.34 & 3.01 & 10.77 & 0.13 & 0.32 & 11.00 & 0.36 & 0.37 & 4.07 & 15.43 & 0.09 & 0.19 & 22.00 & 0.39 & 0.39 & 5.37 & 15.60 & 0.05 & 0.14 \\ 
			\hline
			\multicolumn{22}{c}{D. Concentration parameter = 180, n=250}\\
			\hline
			SEMMS & 0.00 & 0.04 & 0.09 & 4.56 & 0.78 & 0.45 & 0.90 & 0.00 & 0.06 & 0.09 & 5.81 & 0.08 & 0.44 & 0.85 & 0.00 & 0.11 & 0.13 & 6.14 & 0.14 & 0.33 & 0.78 \\ 
			SEMMS-F & 0.00 & 0.02 & 0.09 & 4.56 & 0.78 & 0.48 & 0.92 & 0.00 & 0.05 & 0.09 & 5.81 & 0.08 & 0.47 & 0.91 & 0.00 & 0.09 & 0.11 & 6.14 & 0.14 & 0.39 & 0.83 \\ 
			SEMMS-L & 0.00 & 0.01 & 0.09 & 4.56 & 0.78 & 0.48 & 0.92 & 0.00 & 0.04 & 0.09 & 5.81 & 0.08 & 0.47 & 0.91 & 0.00 & 0.09 & 0.11 & 6.14 & 0.14 & 0.40 & 0.84 \\ 
			Lasso(CV) & 0.00 & 0.20 & 0.20 & 4.91 & 13.38 & 0.15 & 0.36 & 0.00 & 0.23 & 0.23 & 8.93 & 17.28 & 0.09 & 0.25 & 0.00 & 0.29 & 0.29 & 14.47 & 24.43 & 0.04 & 0.10 \\ 
			Lasso(CV)-F & 0.00 & 0.13 & 0.15 & 4.91 & 13.38 & 0.24 & 0.61 & 0.00 & 0.16 & 0.17 & 8.93 & 17.28 & 0.19 & 0.51 & 0.00 & 0.21 & 0.21 & 14.47 & 24.43 & 0.11 & 0.29 \\ 
			Lasso(CV)-L & 0.00 & 0.13 & 0.15 & 4.91 & 13.38 & 0.25 & 0.61 & 0.00 & 0.16 & 0.17 & 8.93 & 17.28 & 0.19 & 0.52 & 0.00 & 0.21 & 0.21 & 14.47 & 24.43 & 0.11 & 0.32 \\ 
			\hline
		\end{tabular}
	\end{adjustbox}
	\clearpage
\end{table}

\begin{table}[ht]
	\clearpage
	\thispagestyle{empty}
	\centering
	\setlength\tabcolsep{2.5pt}  
	\begin{adjustbox}{ addcode=
			{\begin{minipage}{\width}}
				{\caption{Results are based on 500  replications per simulation. 500 mediators are used for the Multiple M case and 500 putative variables are used for the Multiple X case.}\label{medisim}
			\end{minipage}},
			}
\resizebox{\textwidth}{!}{\begin{tabular}{l|rr|rr|rr|rr|rr|rr|rr|rr}
  &\multicolumn{2}{c}{N(0)} & \multicolumn{2}{c}{TP}  & \multicolumn{2}{c}{FP} &
   \multicolumn{2}{c}{$b\ne0$} & \multicolumn{2}{c}{$c^\prime\ne0$}  & 
   \multicolumn{2}{c}{bias($\beta_2$)} & \multicolumn{2}{c}{MAD($\beta_2$)} & \multicolumn{2}{c}{CP($\beta_2$)}\\
 & S & L & S & L & S & L & S & L & S & L & S & L & S & L & S & L \\ 
  \hline
  Multiple M
&    0 &    0 &    1 & 1.000 &    0 & 0.662 &    1 & 1.000 & 0.054 & 0.052 & 0.007 & 0.005 & 0.084 & 0.088 & 0.944 & 0.944 \\  
   $\beta_1=3$, 
 &    1 &    0 &   10 & 2.108 &    0 & 0.630 &    1 & 1.000 & 0.054 & 0.058 & 0.007 & 0.004 & 0.084 & 0.111 & 0.944 & 0.940 \\
   $\beta_2=1$
 &    2 &   19 &   10 & 4.536 &    0 & 0.518 &    1 & 0.898 & 0.054 & 0.054 & 0.007 & 0.006 & 0.084 & 0.238 & 0.944 & 0.954 \\
&    4 &   24 &   10 & 4.630 &    0 & 0.544 &    1 & 0.899 & 0.054 & 0.053 & 0.007 & 0.011 & 0.084 & 0.224 & 0.944 & 0.954 \\ 
  \hline
  Multiple X
&    0 &    0 & 1.000 & 1.000 &    0 & 0.606 &    1 &    1 & 0.054 & 0.054 & 0.007 & 0.006 & 0.084 & 0.084 & 0.944 & 0.946 \\ 
   $\beta_1=3$, 
 &    0 &    0 & 9.876 & 1.632 &    0 & 0.572 &    1 &    1 & 0.054 & 0.056 & 0.007 & 0.005 & 0.084 & 0.084 & 0.944 & 0.950 \\ 
    $\beta_2=1$
 &    0 &    0 & 10.000 & 3.710 &    0 & 0.494 &    1 &    1 & 0.054 & 0.076 & 0.007 & 0.008 & 0.084 & 0.085 & 0.944 & 0.950 \\ 
&    0 &    0 & 10.000 & 3.742 &    0 & 0.494 &    1 &    1 & 0.054 & 0.052 & 0.007 & 0.005 & 0.084 & 0.087 & 0.944 & 0.934 \\
   \hline
   \hline
  Multiple M
&    0 &    0 &    1 & 1.000 & 0.002 & 6.548 &    1 & 1.000 & 0.058 & 0.054 & 0.007 & 0.003 & 0.084 & 0.102 & 0.944 & 0.950 \\ 
   $\beta_1=\sqrt{3}$, 
 &    1 &    0 &   10 & 2.058 & 0.002 & 6.262 &    1 & 1.000 & 0.058 & 0.058 & 0.007 & 0.004 & 0.084 & 0.126 & 0.944 & 0.942 \\ 
    $\beta_2=\sqrt{3}$
 &    5 &   23 &   10 & 4.549 & 0.002 & 5.633 &    1 & 0.998 & 0.057 & 0.048 & 0.007 & -0.000 & 0.084 & 0.242 & 0.943 & 0.941 \\
 &    3 &   30 &   10 & 4.568 & 0.002 & 5.679 &    1 & 1.000 & 0.058 & 0.043 & 0.007 & 0.015 & 0.084 & 0.236 & 0.944 & 0.960 \\ 
  \hline
  Multiple X
&    0 &    0 & 1.000 & 1.000 &    0 & 6.700 &    1 &    1 & 0.058 & 0.048 & 0.007 & 0.006 & 0.084 & 0.102 & 0.944 & 0.948 \\ 
   $\beta_1=\sqrt{3}$, 
 &    0 &    0 & 9.876 & 1.746 &    0 & 6.194 &    1 &    1 & 0.058 & 0.054 & 0.007 & 0.004 & 0.084 & 0.103 & 0.944 & 0.948 \\ 
   $\beta_2=\sqrt{3}$
 &    0 &    0 & 10.000 & 4.068 &    0 & 5.322 &    1 &    1 & 0.058 & 0.068 & 0.007 & 0.009 & 0.084 & 0.101 & 0.944 & 0.956 \\
&    0 &    0 & 10.000 & 4.152 &    0 & 5.650 &    1 &    1 & 0.058 & 0.036 & 0.007 & 0.006 & 0.084 & 0.103 & 0.944 & 0.940 \\ 
   \hline
  \hline
  Multiple M
&    0 &    0 & 1.000 & 1.000 & 0.002 & 9.692 &    1 &    1 & 0.070 & 0.044 & 0.007 & 0.009 & 0.084 & 0.109 & 0.944 & 0.944 \\ 
   $\beta_1=1$,
 &    2 &    0 & 9.972 & 1.880 & 0.002 & 10.004 &    1 &    1 & 0.070 & 0.042 & 0.007 & 0.005 & 0.084 & 0.126 & 0.944 & 0.950 \\
    $\beta_2=3$ 
    &   13 &   21 & 10.000 & 4.167 & 0.002 & 8.962 &    1 &    1 & 0.068 & 0.063 & 0.008 & 0.007 & 0.084 & 0.246 & 0.945 & 0.933 \\
 &    9 &   36 & 10.000 & 4.188 & 0.002 & 8.862 &    1 &    1 & 0.067 & 0.056 & 0.007 & -0.001 & 0.083 & 0.239 & 0.945 & 0.942 \\
  \hline
  Multiple X
&    0 &    0 & 1.000 & 1.000 & 0.006 & 10.202 &    1 &    1 & 0.070 & 0.044 & 0.007 & 0.003 & 0.084 & 0.120 & 0.944 & 0.952 \\ 
   $\beta_1=1$, 
&    1 &    0 & 9.876 & 1.778 & 0.008 & 9.168 &    1 &    1 & 0.070 & 0.058 & 0.007 & -0.001 & 0.084 & 0.116 & 0.944 & 0.952 \\ 
   $\beta_2=3$ 
   &    3 &    8 & 10.000 & 4.093 & 0.008 & 8.307 &    1 &    1 & 0.070 & 0.059 & 0.007 & 0.009 & 0.084 & 0.115 & 0.944 & 0.947 \\
 &    7 &    7 & 10.000 & 4.164 & 0.008 & 8.592 &    1 &    1 & 0.071 & 0.047 & 0.007 & 0.003 & 0.084 & 0.117 & 0.943 & 0.943 \\
   \hline
\end{tabular}}
	\end{adjustbox}
	\clearpage
\end{table}

\end{document}